\documentclass[prd,tightenlines,superscriptaddress,nofootinbib]{revtex4}
\usepackage{amssymb,latexsym}
\usepackage{amsmath,amsbsy,bbm}
\usepackage{epsfig,bm}
\usepackage{graphicx,comment}
\usepackage{color}
\begin{document} 

\title{Evidence of anomalous Curie constants for nonmagnetic impurities in a critical 2-dimensional $JQ_3$ model on the honeycomb lattice}

\author{L.-W. Huang}
\affiliation{Department of Physics, National Taiwan Normal University, 
  88, Sec.4, Ting-Chou Rd., Taipei 116, Taiwan}
\author{J.-H. Peng}
\affiliation{Department of Physics, National Taiwan Normal University,
88, Sec.4, Ting-Chou Rd., Taipei 116, Taiwan}
\author{F.-J. Jiang}
\affiliation{Department of Physics, National Taiwan Normal University,
88, Sec.4, Ting-Chou Rd., Taipei 116, Taiwan}
\email[]{fjjiang@ntnu.edu.tw}
\vspace{-2cm}
  
\begin{abstract}
The Curie constants $C^{*} = \lim_{T\rightarrow 0}T\chi_{\text{imp}}$ of a spin-1/2 and a spin-1 impurities are calculated using the quantum Monte Carlo simulations. Here the impurity
susceptibility $\chi_{\text{imp}}$ is the difference between
the uniform susceptibilities with and without the impurity and $T$ is the temperature. Moreover,
	the two-dimensional quantum $JQ_3$ model on the 
	honeycomb lattice is considered as the host system. 
	$T\chi_{\text{imp}}$ as a function $T$ is investigated in great detail.
        Remarkably, our data indicate strongly that $C^* > 0.25$ and $C^* > 2/3$ for the
        spin-1/2 and the spin-1 impurities, respectively.
        In particular, although no definite conclusion is obtained due to finite-size effects, 
	we find the Curie constant associated with a spin-1/2 impurity likely
	converges to a value greater than the established result of 0.262(2) 
	in the literature. The outcomes reached here
        provide certain evidence that fractional impurity spin is observed for the studied $JQ_3$ system.

\end{abstract}

\maketitle
\vskip-0.5cm

\section{Introduction}
Effects of impurities in antiferromagnets have triggered theoretical interest.
This is because the responses of the host systems due to the 
impurities provide
an important information for understanding the associated bulk properties \cite{Aha88,Bul89,Mao91,Egg92,Sch94,Iag95,San97,Mar97,Sus00,Nis00,Nis001,Vaj02}. 
Doped holes in an antiferromagnet is one kind of impurities. When holes are doped into cuprate materials, 
at low concentration the mobility of these doped holes is very low. Hence, the doped holes are localized in the 
host materials. As a result, static impurities are relevant for investigating the weakening of antiferromagnetic
order upon doping charge carriers into the cuprate insulators. 

Recently, the magnetic response of the bulk system to a localized spin-$S$
impurity has been investigated in great detail using analytic field theory method \cite{Sac99,Voj00,Sac03}. Since then several Monte Carlo studies
of impurities in antiferromagnets have been carried out \cite{Sac01,Tro02,Hog03,Hog04,Hog071}, and some outcomes from the relevant Monte Carlo results agree
nicely with the predictions of 
Refs.~\cite{Sac99,Voj00,Sac03}.

At a quantum-critical spin-1/2 antiferromagnet, the impurity susceptibility $\chi_{\text{imp}}$, which is defined as the difference of the 
uniform susceptibilities with and without the spin-$S$ impurity, has the Curie form $T\chi_{\text{imp}} \rightarrow C^{*}$ when $T \rightarrow 0$. Here $T$
is the temperature and
$C^{*}$ is the Curie constant taking the expression $C^{*} = \tilde{S}(\tilde{S}+1)/3$ with $\tilde{S}$ being some constant.  
The Curie constant $C^{*}$ is argued to satisfy the inequality $S^2/3< C^{*} < S(S+1)/3$ which
can be interpreted as a fractional impurity spin $\tilde{S} \neq S$. However, based on the Green's function theory,
it is shown that $\tilde{S} = S$ \cite{Sus03}. 

The attempts so far to determine the numerical value of $C^{*}$ through Monte Carlo simulations, 
hence to better understand the conjecture of a fractional impurity spin, 
are done for two-dimensional (2D) symmetric and incomplete bilayer quantum Heisenberg models \cite{Tro02,Hog07}. For the symmetric bilayer model, 
while the numerical Monte Carlo
results in Ref.~\cite{Tro02} seem to support the fact that the fractionalization is either very small or is not present, the data points
shown in Ref.~\cite{Hog07} do not lead to a definitive answer.
The most noticeable finding among these Monte Carlo studies related to the determination of $C^{*}$
is the report of an anomalous Curie constant from simulating 
the incomplete bilayer model \cite{Hog07}. Specifically, for incomplete quantum bilayer model, a convincing 
numerical evidence suggests that $C^{*} = 0.262(2)$,
which falls outside the conjectured value $ S^2/3 \le C^{*} \le S(S+1)/3$ for a spin-$1/2$ impurity \cite{Sac99,Voj00,Hog07}. 
The result of $C^{*} = 0.262(2)$ also disagrees with the value determined by
the Green's function theory calculation \cite{Sus03}. Finally, for a spin-1 impurity, the related Curie constant $C^*$ is found to
be consistent with $C^* = 2/3$ which implies the lack of fractionalization \cite{Hog07}.

To shed some light on the existing discrepancy regarding
whether $C^{*}$ is in agreement with the scenario of a fractional impurity spin, particularly to examine if $C^{*}$ of a spin-1/2 impurity has
an anomalous result of $0.262(2)$ as reported in Ref.~\cite{Hog07},
in this study we perform a large-scale quantum Monte Carlo simulations (QMC) to calculate the values of $C^{*}$ related to the 2D quantum
$JQ_3$ models (defined later) on the honeycomb lattice.

The host system chosen here is the 2D quantum $JQ_3$ model on the honeycomb lattice \cite{Puj13,Pen22}, and the 
corresponding spin-1/2 impurity system is obtained by the removal of one single spin from the host system.
Moreover, a spin-1 impurity is created by considering the couplings touching a particular quantum spin to be
ferromagnetic, while the other couplings remain antiferromagnetic.
The QMC are carried out at the associated critical point of the $JQ_3$ model.
The $JQ_3$ model is considered here due to its exotic criticality \cite{Sen040,Sen041}. 
Apart from determining the values of $C^{*}$, this model is considered in our investigation because the associated phases transition is
unusual and exotic. Specifically, for the $JQ_3$ model, although the breaking
and restoration of symmetries 
for the two sides of the related critical point are
different, yet a genetic second order phase transition may occur \cite{Sen040,Sen041}
(Such an exotic criticality is termed deconfined quantum criticality (DQC) in the literature).
Consequently, by studying the response of the $JQ_3$ model
to a nonmagnetic impurity may reveal certain features of the nature of the corresponding critical theory.

Based on the obtained data, for intermediate temperatures and large lattices we find $T\chi_{\text{imp}} > 0.25$ for the system with a
spin-1/2 impurity. This is consistent with the known outcomes in the literature. Remarkably, our results indicate the
quantity $T\chi_{\text{imp}}$ can reach
a value as large as 0.3. This number 0.3 is statistically different than the result of 0.262(2) obtained in Ref.~\cite{Hog07}.
In addition, we observe that for the case of spin-1 impurity, the associated $C^*$ is likely larger than $2/3$ as well.
Our investigation presented here implies fractional impurity spin(s) is (are) observed in the considered quantum $JQ_3$ model.
Whether the difference between the results obtained here and that in previous works is due to the fact
that the considered host models are different requires further exploration.

This paper is organized as follows. First, after an introduction,
the spin-1/2 host systems, namely the $JQ_3$ models on the honeycomb lattice and the relevant observables studied in this work are briefly described.
Then we present our numerical results in detail. In particular, the evidence to support the scenario described above is demonstrated.
Finally, a section concludes our study.

\begin{figure}
 \begin{center}
    \vskip-0.2cm
\includegraphics[width=0.4\textwidth]{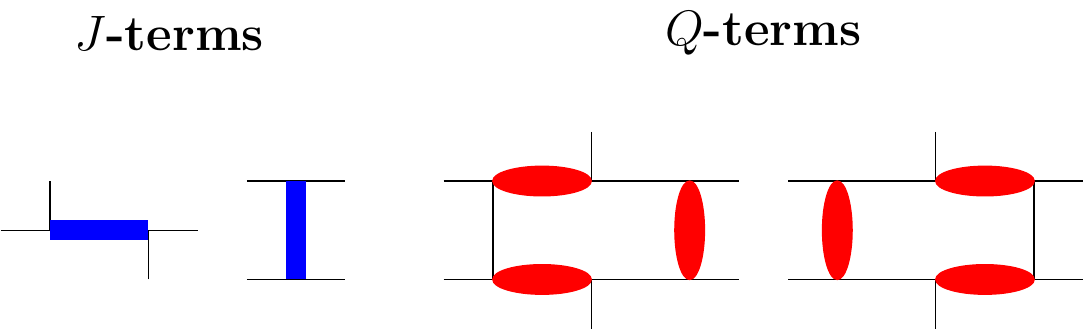}\vskip0.6cm
\end{center}\vskip-0.7cm
  \caption{The host system (quantum $JQ_3$ models) on the
    honeycomb lattice considered in our investigation.}
\label{fig1}
\end{figure}

\section{Microscopic Models and Corresponding Observables}
The host 2D quantum $JQ_3$ model on the honeycomb lattice
we consider in this study is defined by the Hamiltonian
\begin{eqnarray}
\label{hamilton}
H &=& J\sum_{\langle ij \rangle}\vec{S}_i\cdot\vec{S}_j - Q\sum_{\langle ijklmn\rangle}P_{ij}P_{kl}P_{mn},\nonumber \\ 
P_{ij} &=&\frac{1}{4}-\vec{S}_i\cdot\vec{S}_j,
\end{eqnarray}
where in Eq.~(1) $J$ (which is set to be 1 here) and $Q$ are the
couplings for the two-spin and the six-spin interactions, respectively,
$\vec S_{i} $ is the spin-1/2 operator at site $i$, 
$\langle i j \rangle$ denotes a pair of nearest neighbor sites $i$ and $j$, 
and $P_{lm}$ is the singlet pair projection
operator between nearest neighbor spins located at sites $l$ and $m$. 
Fig.~\ref{fig1} contains the cartoon representation of the investigated model.
For simplicity, here the $JQ_3$ model
will be named the honeycomb model.

By creating a vacancy
in the host system, one sub-lattice has one more spin
than the other. Hence, the removal of a single spin from the
host system described by Eq.~(\ref{hamilton}) effectively creates
a spin-1/2 (nonmagnetic) impurity in the system.
In addition, a spin-1 impurity is obtained when the couplings
touching a particular quantum spin is set to ferromagnetic while
the other couplings remain antiferromagnetic.

To calculate the Curie constants,
the impurity susceptibility $\chi_{\text{imp}}$, which is the difference
between the (uniform) susceptibilities of the host system with and
without an impurity, is measured in our study. 
Specifically, $C^{*}$ is given by
\begin{equation}
C^{*} = \lim_{T\rightarrow 0}T\chi_{\text{imp}},
\end{equation}
where $T$ is the temperature and $\chi_{\text{imp}} = \chi_{1} - \chi_{0}$.
Here $\chi_{1} $ and $\chi_{0}$ are the susceptibilities of the system
with and without an impurity, respectively, and are defined by 
\begin{equation}
\chi_1 =\frac{J}{T}\left(\sum_{i=1}^{M_1}S_i^{z}\right)^2,\,\,\chi_0 = \frac{J}{T}\left(\sum_{i=1}^{M_0}S_i^{z}\right)^2
.\end{equation}
Here for a system with linear box size $L$, $M_0 = L^2$ and $M_1 = L^2-1$ or $M_1 = L^2$ for the spin-1/2
and spin-1 impurities, respectively.

\section{Numerical Results}
To determine the associated Curie constant $C^{*}$ of an impurity in a host system, extensive simulations at the
related critical point is needed.
The estimated critical point $(Q/J)_c$ for the considered $JQ_3$ model is given by
$(Q/J)_c \sim 1.1396$ \cite{Puj13}.
Hence, we have carried out large-scale quantum Monte Carlo calculations (QMCs) with $Q/J = 1.1396$
using the stochastistic series expansion (SSE) 
algorithm with very efficient operator loop update \cite{San99,Syl02,San10}.
In addition, the calculations are conducted with several box sizes $L$ at various temperatures $T$, and periodic boundary
conditions are implemented for both the spatial directions. With the conventional labels of spins (and their spatial coordinates),
for a $L$ by $L$ square lattice the particular spin associated with the calculations is chosen to be last one. Finally,
The errors of $\chi_{\text{imp}}$ are determined by either the standard method or a boostrap procedure.

\begin{figure}
  \vskip0.3cm
\begin{center}
\includegraphics[width=0.45\textwidth]{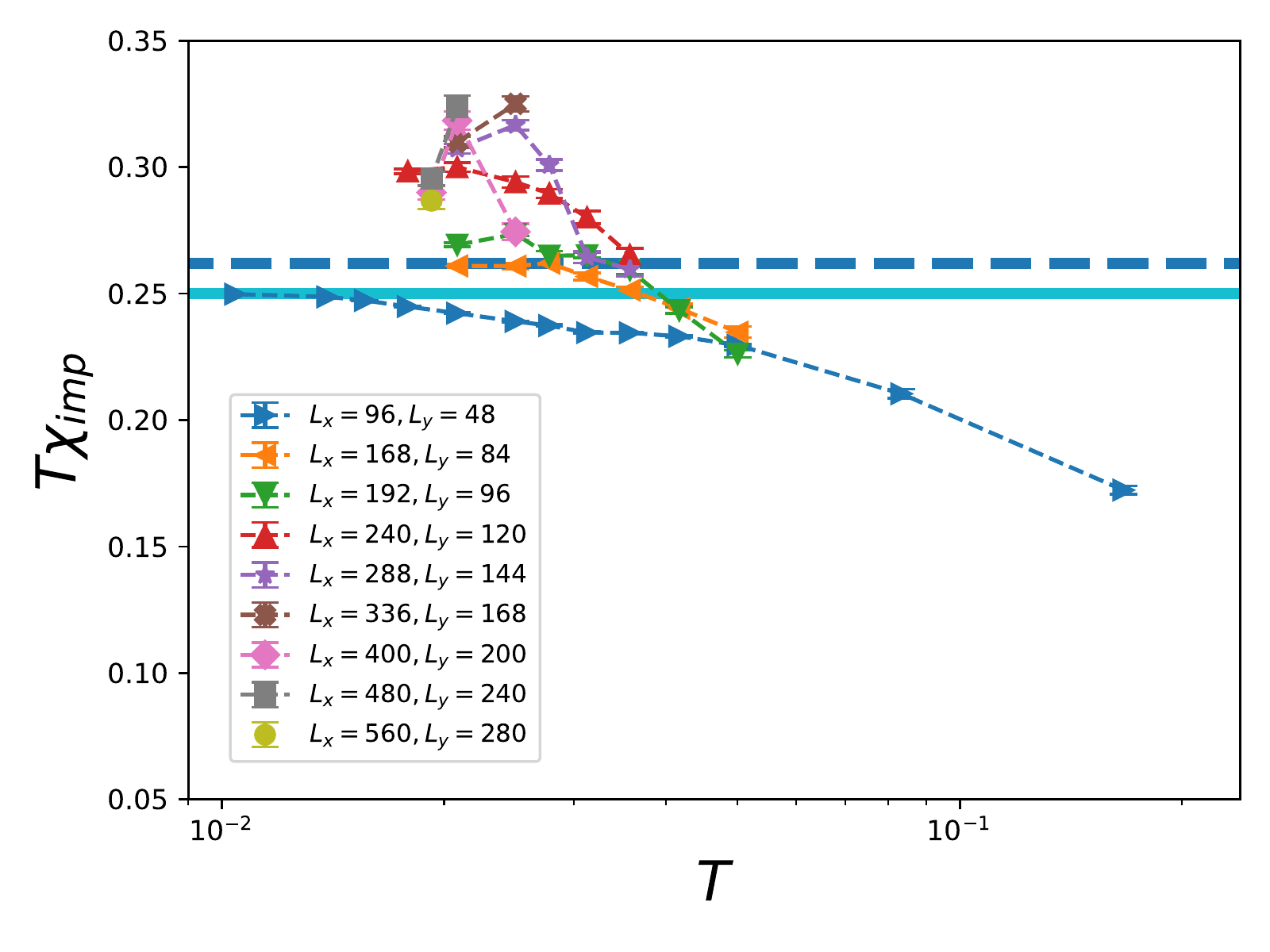}
\end{center}
\vskip-1.0cm
\caption{$T\chi_{\text{imp}}$ as functions of $T$ for various box sizes. The horizontal dashed and solid lines represent 0.262 and 0.25, respectively.}
\label{fig2}
\end{figure}

For the case of spin-1/2 impurity, $T\chi_{\text{imp}}$ as functions of $T$ for various honeycomb lattices are demonstrated in fig.~\ref{fig2}.
Interestingly, for larger system sizes,
values of this quantity are greater than 0.25 (the horizontal solid line in fig.~\ref{fig2}).
Moreover, many of the obtained outcomes are even larger than 0.262 (the horizontal dashed line in fig.~\ref{fig2}),
which was
the bulk $C^{\star}$ determined in Ref.~\cite{Hog07}. Although no definite conclusion can be drawn from fig.~\ref{fig2}
due to the highly non-monotonic $L$
dependence of $T\chi_{\text{imp}}$ at any fixed $T$ (Hence, an extrapolated fit cannot be conducted),
it is of high probability that the $C^{*}$ corresponding to the considered $JQ_3$ model fulfills $C^{*} \gtrsim 0.3$.

We would like to point out that due to finite-size effects, for a fixed $L$ the quantity $T\chi_{\text{imp}}$
associated with a spin-1/2 (spin-1) impurity will approach 1/4 (2/3) at low temperatures.
As a result, $C^*$ (lim$_{T\rightarrow 0}T\chi_{\text{imp}}$) should be
calculated by firstly carrying out the extrapolation in $L$ at every fixed $T$, and then the obtained outcomes
are used to conduct the $T$-extrapolation. Because of the highly non-monotonic $L$
dependence of $T\chi_{\text{imp}}$ at any fixed $T$ as can be seen in fig.~\ref{fig2}, such a procedure cannot be done here.
Still, the trend of $T\chi_{\text{imp}}$ data points as functions of $L$ and $T$ indicates the high possibility of
$C^* \gtrsim 0.3$ for a spin-1/2 impurity.

It should be emphasized that in Ref.~\cite{Hog07}, a size convergence criterion is used to determine the bulk value of
$T\chi_{\text{imp}}$ for every considered $T$. Specifically, a quantity is considered to be size converged if its values
at size $L$ and size $2L$ agree within statistical uncertainties. One has to be extremely careful when this criterion is employed.
This is because of the nonmonotonic behavior when $T\chi_{\text{imp}}$ is considered as a function of $L$ at a fixed $T$. Indeed,
for $T=0.025$, while the values of $T\chi_{\text{imp}}$ for $L_1 = 400$ and $L_1 = 192$ agree with each other, this quantity associated
with $L_1 = 288$ differs from those of $L_1 = 400$ and $L_1=192$.

\begin{figure}
  \vskip0.3cm
\begin{center}
\includegraphics[width=0.45\textwidth]{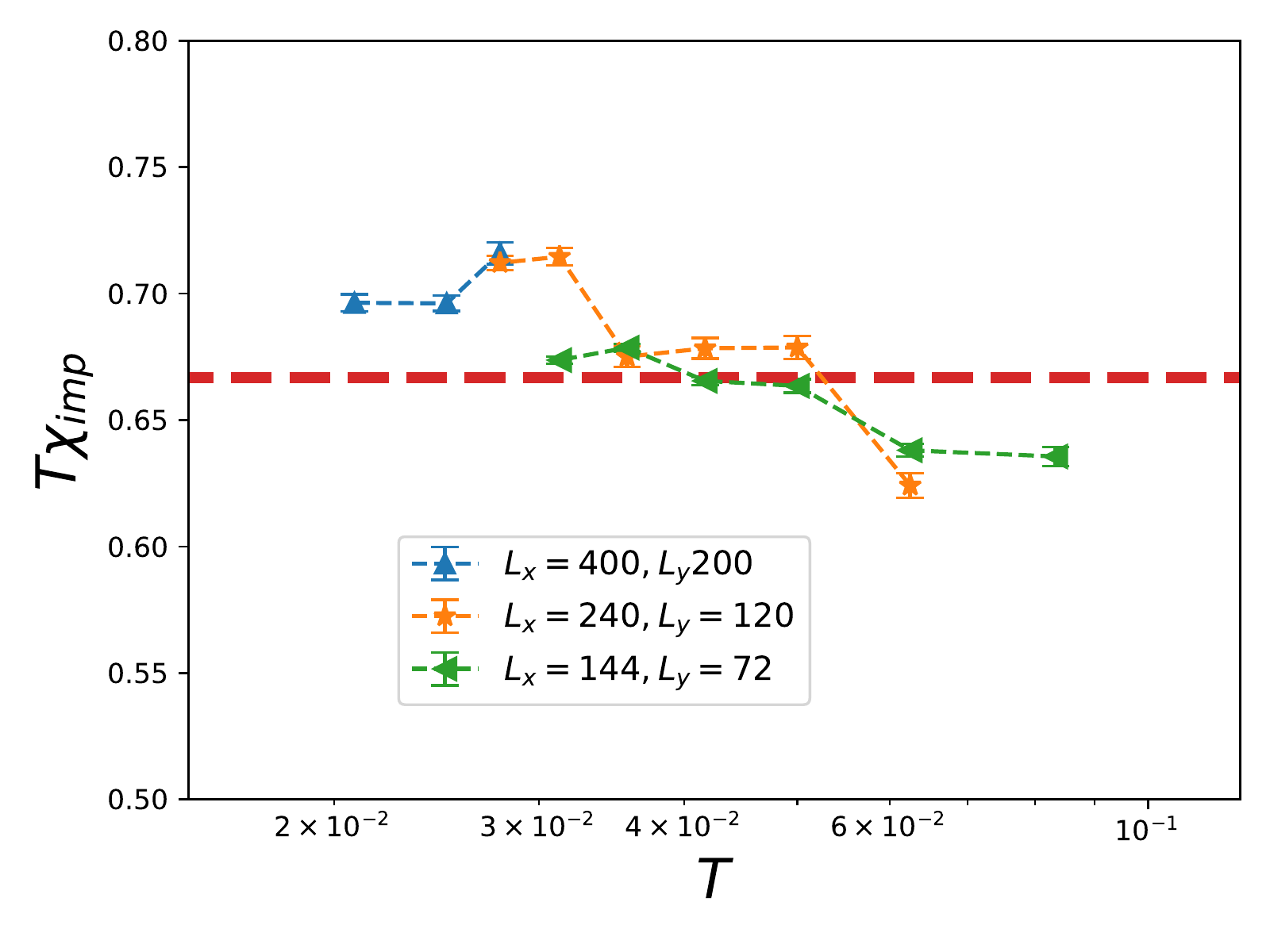}
\end{center}
\vskip-1.0cm
\caption{$T\chi_{\text{imp}}$ as functions of $T$ for various box sizes. The horizontal dashed line represents 2/3.}
\label{fig3}
\end{figure}

After demonstrating the outcomes of a spin-1/2 impurity, we turn to the case of spin-1 impurity. $T\chi_{\text{imp}}$ as functions of
three honeycomb lattices and various $T$ are shown in fig.~\ref{fig3}. Similar to the results of spin-1/2 impurity, many data points
presented in fig.~\ref{fig3} have magnitude larger than 2/3. In particular, unlike the claim made by a previous investigation \cite{Hog07}, our data suggest that the Curie constant $C^*$ associated
with a spin-1 impurity likely has a value greater than 2/3.

\section{Discussions and Conclusions}
In this study, we have performed large-scale QMCs to calculate
the Curie constants $C^{*}$ associated with a spin-1/2 and a spin-1 impurities.
In particular, the host system is the 2D quantum $JQ_3$ model on
the honeycomb lattice.
The related spin-1/2 impurity system is obtained by the removal of
a single spin in the host system. Moreover, the spin-1 impurity
system is reached by considering the couplings touching a particular
quantum spin to be ferromagnetic.

Although no definite conclusions are obtained due to finite-size effects, 
	we find that the $C^*$ of a spin-1/2 impurity likely
	converges to a value greater than the established result of 0.262(2) 
	in the literature. For the case of spin-1 impurity, there is a great chance
        that the associated $C^*$ fulfills $C^* > 2/3$. The outcomes reached here also
        provide certain evidence that fractional impurity spins are observed for the studied $JQ_3$ system.

        We would like to point out that the results presented here use huge amount of computing time. For instance, for the largest
        lattice with the lowest temperature considered in this study alone takes more than $4 \times 10^5$ core-hours using xeon E5-2620v2 (and opteron 6344). 
        
The discrepancy regarding the outcomes obtained here and that established in the literature may be
explained by the fact that the host models considered in this study and that in Ref.~\cite{Hog07} are different.
To better understand the properties of $C^{*}$, 
it will be extremely interesting to calculate the $C^{*}$
associated with other host systems such as the $JQ_3$ model on the square lattice as the one studied in Ref.~\cite{Lou09},
or the 2D dimerized quantum antiferromagnets with staggered- or plaquette-dimer pattern \cite{Wen08,Wen09,Jia09,Fri11,Jia12}. 
Finally, due to the potential
experimental realization of DQC \cite{Cui} in the material SrCu$_2$(BO$_3$)$_2$, the results presented in our
study may be verified in future relevant experiments.

\section{Acknowledgements}
\vskip-0.5cm

The first two authors contribute equally to this project. Partial support from the Ministry of Science and Technology of Taiwan is acknowledged.

\end{document}